\documentclass[aps,prl, twocolumn, floatfix]{revtex4-1}
\pdfoutput = 1
\usepackage{graphicx}
\usepackage{hyperref}
\usepackage{amsmath}
\usepackage{color}
\usepackage{dcolumn}
\usepackage{bm}
\usepackage{float}
\usepackage[normalem]{ulem}
\usepackage{braket}
\usepackage[subrefformat=parens,labelformat=parens,caption=false]{subfig}
\usepackage{bbm}
\usepackage[dvipsnames]{xcolor}
\usepackage{esint}
\usepackage{lineno}
\usepackage{verbatim}
\usepackage{bm}
\usepackage[export]{adjustbox}

\usepackage{soul}

\begin{document}
\title{Noise, not squeezing, boosts synchronization in the deep quantum regime}

\author{W. K. Mok$^{1}$, L.~C.\ Kwek$^{1,2,3,4}$, H.\ Heimonen$^{1}$}
\affiliation{$^{1}$Centre for Quantum Technologies, National University of Singapore, 3 Science Drive 2, Singapore 117543}
\affiliation{$^{2}$Institute of Advanced Studies, Nanyang Technological University, 60 Nanyang View, Singapore 639673}
\affiliation{$^{3}$National Institute of Education, Nanyang Technological University, 1 Nanyang Walk, Singapore 637616}
\affiliation{$^{4}$MajuLab, CNRS-UNS-NUS-NTU International Joint Research Unit, UMI 3654, Singapore}

\begin{abstract}
Synchronization occurs ubiquitously in nature. The van der Pol oscillator has been a favorite model to investigate synchronization. Here we study the oscillator in the deep quantum regime, where nonclassical effects dominate the dynamics. Our results show: (i) squeezed driving loses its effect, (ii) noise boosts synchronization, (iii) synchronization is bounded, and (iv) the limit-cycle is insensitive to strong driving. We propose a synchronization measure and analytically calculate it. These results reflect intrinsic differences between synchronization in the quantum and deep quantum regimes.

\end{abstract}

\date{\today}
\maketitle

{\em Introduction.}-- From coupled metronomes\citep{strogatz2004sync,balanov2009synchronization,pikovsky2003synchronization}, to brain activity\citep{cabral2011role}, and jet-lag\citep{kantermann2007human}, synchronization is used to explain countless phenomena in nature. In practice, many of these processes are noisy, with random noise influencing the dynamical systems. This has given rise to many noise-enhanced, and noise-enabled processes, such as signal amplification \citep{chia2019phase} and stochastic resonance\citep{gammaitoni1998stochastic, benzi1981mechanism}. Studies of synchronization have been taken to the quantum domain and various quantum systems with limit cycles have been studied in recent years, including the weakly nonlinear quantum van der Pol (qvdP) oscillator\citep{lee2013quantum, walter2014quantum,walter2015quantum,lorch2017quantum,lorch2016genuine,morgan2015oscillation,chia2017quantum,bastidas2015quantum}, optomechanical systems\citep{heinrich2011collective, Florian2013,weiss2016noise,weiss2017quantum,bagheri2013photonic,amitai2017synchronization}, and low-dimensional systems\citep{roulet2018synchronizing,koppenhofer2019optimal, koppenhofer2019quantum,zhang2019quantum,parra2019synchronization,ilves2019bath}. Interest in lies in understanding the difference between quantum and classical synchronization \citep{sonar2018squeezing,lorch2016genuine,lorch2017quantum}, realizing a quantum self-sustained oscillator \citep{laskar2019observation,koppenhofer2019quantum}, and using quantum synchronization for tasks such as operating heat engines\citep{jaseem2020quantum}, and generating entanglement\citep{roulet2018quantum, witthaut2017classical}. 

In this Letter we push the qvdP into the deep quantum regime (characterised by $\gamma_2/\gamma_2 \gtrsim 10$), and the deep quantum limit ($\gamma_2 / \gamma_1 \to \infty$). Experimentally, single photon losses cannot be removed from the oscillator. Yet, we find that it can boost synchronization in this regime. We also find that the amount of synchronization to an external drive is capped in the deep quantum regime. Moreover, we discover that the effect of a squeezing drive vanishes and we obtain an upper bound to the amount of synchronization for a harmonically driven qvdP. The oscillator is also shown to be extraordinarily robust against perturbation from driving.

{\em Model.}-- To study quantum synchronization, we require a self-sustained (limit-cycle) oscillator. The most commonly studied model is the quantum van der Pol oscillator in the weakly nonlinear regime \citep{lee2013quantum,walter2014quantum}, also known as the Stuart-Landau oscillator \citep{kuramoto2012chemical}. We add linear damping to the standard qvdP, fully described by the master equation
\begin{equation}
\label{eq:ME}
\dot{\rho} = -i [H, \rho] + \gamma_1 \mathcal{D}[a^\dag]\rho + \gamma_2 \mathcal{D}[a^2]\rho + \kappa \mathcal{D}[a]\rho
	\end{equation} 
with the Hamiltonian given by
	\begin{equation}
H = \delta a^\dag a + \Omega (a + a^\dag) + \eta (a^2 + a^{\dag 2})
	\end{equation}
in the rotating frame with the harmonic and squeeze drives, using the rotating-wave approximation. The qvdP is driven by a harmonic drive of strength $\Omega$ and a squeezing Hamiltonian of strength $\eta$. The oscillator detuning is $\delta = \omega_0 - \omega_d$, where $\omega_0$ is the natural frequency of the oscillator and $\omega_d$ is the driving frequency. The limit cycle is maintained by three incoherent processes: single-photon pumping with the pumping rate $\gamma_1$, two-photon loss with decay rate $\gamma_2$, and a single-photon loss with decay rate $\kappa$. The nonlinear, two photon, loss is necessary for the limit cycle to exist. Consequently, only single photon loss is regarded as noise. These incoherent processes do not imprint any phase preference on the oscillator, leaving the phase free and enabling the oscillator to synchronize.

{\em Analytics.}-- The optical phase is a periodic variable in the interval $[0,2\pi)$. Therefore it is natural to use directional statistics \citep{mardia2009directional} to study its properties. To measure phase synchronization we use the mean resultant length (MRL) of a circular distribution $S = \sqrt{\braket{\sin{\phi}}^2+\braket{\cos{\phi}}^2} = |{\braket{e^{i\phi}}}|$, where the expectation value is taken over the probability distribution. The measure is closely related to the circular variance\citep{mardia2009directional}, is used to measure synchronization for classical noisy systems \citep{allefeld2004testing}, and is also famously the order parameter of the Kuramoto model\citep{kuramoto2012chemical}. We choose the MRL as our synchronization measure for its suitable properties: (i) It takes the value 0 for an un-synchronised state, and the value 1 for a perfectly synchronised state where the phase-distribution is a delta function, so it has a clear numerical meaning, unlike unbounded measures. (ii) The MRL has a natural counterpart as the expectation value of a quantum operator, given by $S= |\braket{e^{i \hat{\phi}}}| = |\textrm{Tr}(e^{i\hat{\phi}} \rho )|$, where $\hat{\phi}$ is the phase operator defined by Pegg and Barnett\citep{pegg1989phase}. (iii) Having a quantum analogue allows for analytical calculations of the amount of synchronisation in the system. Another feature of synchronization is frequency entrainment. It is quantified by the shifting of the peak of the frequency spectrum.

With pure harmonic driving of the qvdP ($\eta = 0$), we can obtain an analytical approximation for the master equation in the deep quantum regime. To this end, we make the following ansatz for the density matrix in the Fock basis:
	\begin{equation}
\rho_{ss} = \left(\begin{matrix}\rho_{00}&\rho_{01}&0\\\ \rho_{10}&\rho_{11}&0\\0&0& \rho_{22}\end{matrix}\right)
\label{eq:rho}
	\end{equation} 
This amounts to restricting the number of excitations to 2, and neglecting all coherences involving the state $\ket{2}$. The limit-cycle amplitude $N = \braket{a^\dag a} = \rho_{11} + 2 \rho_{22}$ is then easily calculated:
	\begin{equation}
	\label{eq:N_dql}
\lim_{\gamma_2/\gamma_1 \to \infty} N = \frac{\gamma _1 \left(6 \gamma _1 \kappa +9 \gamma _1^2+4 \delta ^2+\kappa ^2+12 \Omega ^2\right)+4 \kappa  \Omega ^2}{\left(3 \gamma _1+\kappa \right) \left(6 \gamma _1 \kappa +9 \gamma _1^2+4 \delta ^2+\kappa ^2+8 \Omega ^2\right)}
	\end{equation}
where the undriven amplitude is given by
	\begin{equation}
N_0 = \frac{\gamma _1 \left(2 \gamma _1+\gamma _2+\kappa \right)}{\gamma _1 \left(3 \gamma _2+\kappa \right)+\kappa  \left(\gamma _2+\kappa \right)+\gamma _1^2}
	\label{eq:N0}
	\end{equation}
which simplifies in the deep quantum limit to
	\begin{equation}
\lim_{\gamma_2 / \gamma_1 \to \infty} N_0 = \frac{\gamma _1}{3 \gamma _1+\kappa }
	\label{eq:N0_dql}
	\end{equation}
As a check for consistency, we note that setting $\kappa = 0$ in Eq (\ref{eq:N0_dql}) gives $N_0 = 1/3$, which agrees with the well known result for the undriven qvdP in the deep quantum limit: $\lim_{\gamma_2 / \gamma_1 \to \infty} \rho_{ss} = \frac{2}{3} \ket{0}\bra{0} + \frac{1}{3} \ket{1} \bra{1}$ \cite{morgan2015oscillation,lee2013quantum}. 

\begin{figure}
\subfloat{%
  \includegraphics[width=\linewidth]{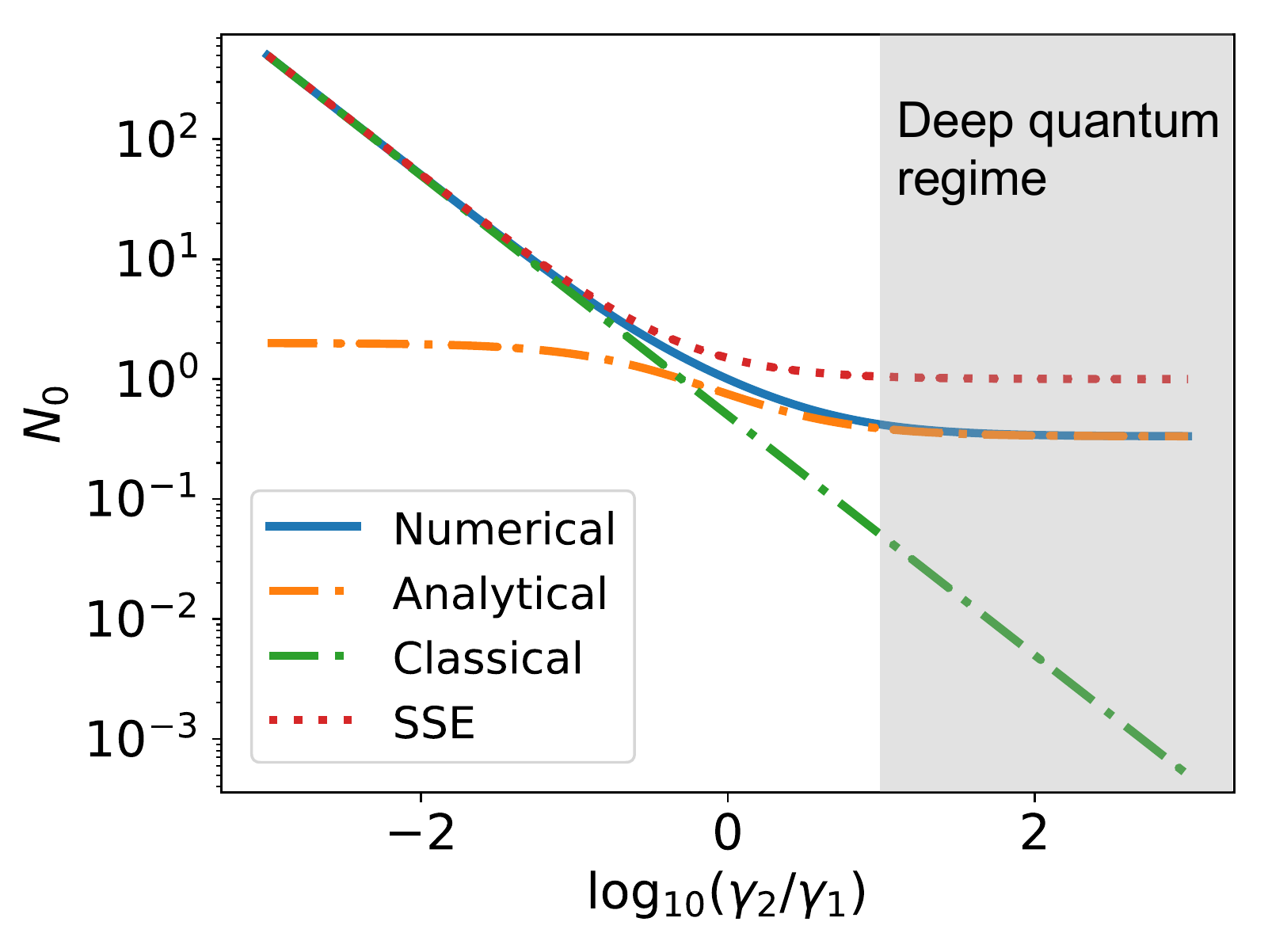}%
}\hfill
	\caption{Amplitude $N_0$ of the undriven noiseless qvdP oscillator. Eq.(\ref{eq:N0}) agrees well with the numerical simulations in the deep quantum regime $\gamma_2/\gamma_1 \gtrsim 10$. The classical curve refers to a mean-field calculation, and SSE refers to a semi-classical system-size expansion.}
	\label{fig:N0_comparison}
\end{figure}

In Fig.\ref{fig:N0_comparison} we compare the noiseless ($\kappa = 0$) oscillator amplitude $N_0$ calculated numerically for various $\gamma_2 / \gamma_1$ with three analytical approximations. The classical calculation refers to a mean-field approximation $\braket{\hat{a}} = \alpha$ that discards all quantum fluctuations. This approximation works well in the large amplitude limit, but breaks down as $\gamma_2 / \gamma_1 \approx 1$. A semi-classical calculation using a system size expansion (SSE) of the equation of motion of the Wigner function \citep{heimonen2020realistic, carmichael2003statistical} produces the expression $N_0 = (\gamma_1 + 2 \gamma_2 - \kappa) / (2\gamma_2)$. The SSE performs better than the classical result in the quantum regime, but it too fails in the deep quantum regime as its limiting value is $N_0 = 1$, not $N_0 = 1/3$. Finally, the calculation in the deep quantum limit in Eq.(\ref{eq:N0}) works well for $\gamma_2 / \gamma_1 \gtrsim 10$. This is the first reason for defining the deep quantum regime to start at the damping ratio 10.
\begin{table}
    \begin{tabular}{ |l | r | l | p{3.5cm} |}
    \hline
    \textbf{Regime} & $\boldsymbol{\gamma_2}/\boldsymbol{\gamma_1}$ & \textbf{Analytical Method} \\ \hline
    Classical limit & 0 & Mean-field \\ \hline
    Semi-classical regime &$ \lesssim 0.1$ &System Size Expansion \\ \hline
    Quantum regime & 1 & None  \\ \hline
    Deep quantum regime & $\gtrsim 10$ & Density matrix ansatz \\ \hline
    Deep quantum limit & $\infty	$ & Density matrix ansatz \\ \hline
    \end{tabular}
    \caption{Classifying the different regimes of the qvdP, based on the analytical methods that work in each.}
    \label{table:regimes}
\end{table}

\begin{figure}
\subfloat[$\gamma_2/\gamma_1 = 100$]{%
  \includegraphics[width=0.499\linewidth]{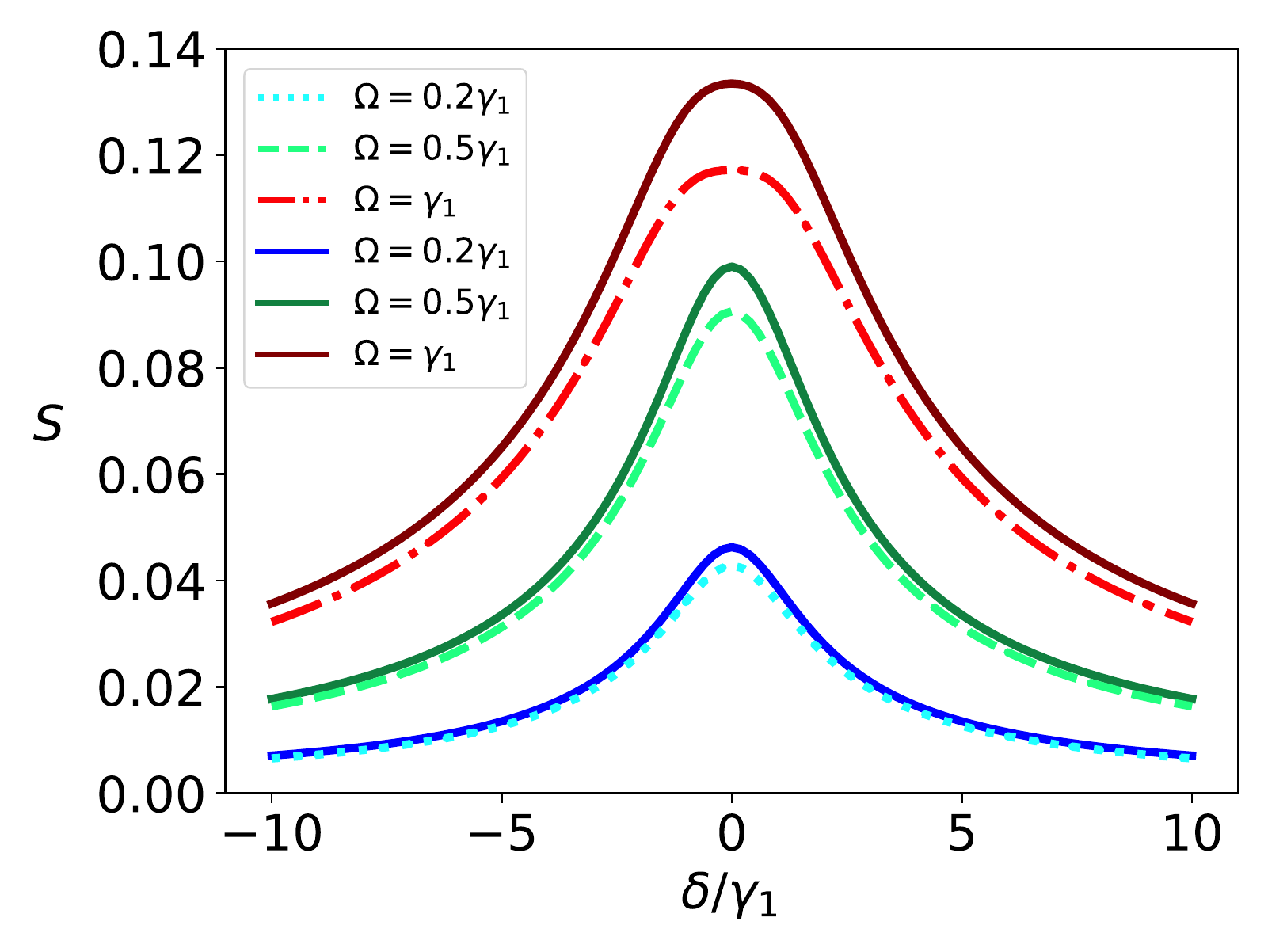}%
  \label{fig:tongue_compare_b=100}%
}\hfill
\subfloat[$\gamma_2/\gamma_1 = 1000$]{%
  \includegraphics[width=0.499\linewidth]{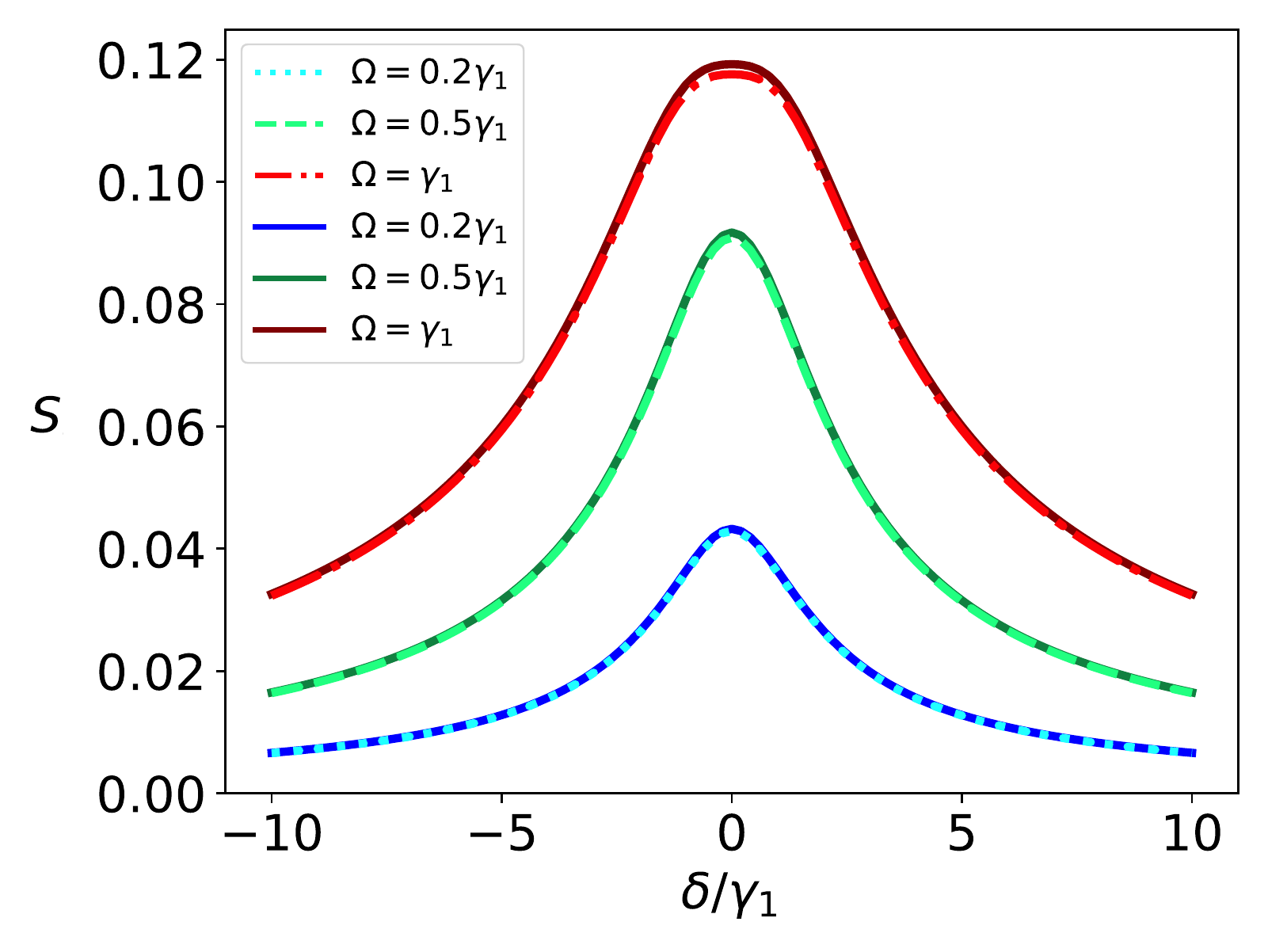}%
  \label{fig:tongue_compare_b=1000}%
}
\hfill
	\caption{Slices of the Arnold tongue. Synchronization measure plotted against detuning for various driving strengths $\Omega/\gamma_1$. The solid lines denote numerical results and the solid lines analytical predictions from Eq.(\ref{eq:sync_dql}).}
	\label{fig:arnold_comparison}
	\end{figure}

The synchronization measure is then identified from Eq.(\ref{eq:rho}) as simply the coherence $|\rho_{01}|$, which can be found by solving for the matrix elements. The full expression is cumbersome, so we present only the steady-state solution in the deep quantum limit.
	\begin{equation}
\lim_{\gamma_2/\gamma_1 \to \infty} S = \frac{2 \Omega  \left(\gamma _1+\kappa \right)\sqrt{(3 \gamma_1 + \kappa)^2 + 4\delta^2}}{\left(3 \gamma _1+\kappa \right) \left(6 \gamma _1 \kappa +9 \gamma _1^2+4 \delta ^2+\kappa ^2+8 \Omega ^2\right)}
	\label{eq:sync_dql}
	\end{equation}
Fig.(\ref{fig:arnold_comparison}) compares the expression to numerical results for different damping ratios $\gamma_2/\gamma_1$, showing the accuracy of the expression in the deep quantum regime. We can also obtain the phase distribution $P(\phi)$ by projecting the qvdP state onto the phase eigenstates: $\ket{\phi} = \sum_n e^{in\phi} \ket{n}$
	\begin{equation}
P(\phi) = \frac{1}{2\pi} \braket{\phi|\rho|\phi} = \frac{1}{2\pi} \sum_{m,n=1}^\infty e^{i(n-m)\phi} \rho_{mn}
	\end{equation}
In the deep quantum limit, we can obtain a simple expression for the phase distribution

	\begin{equation}
	\begin{split}
\lim_{\gamma_2/\gamma_1 \to \infty} P(\phi) =& \frac{1}{2\pi} \bigg[ 1-\frac{4\Omega (\kappa + \gamma_1)}{ 6\gamma_1 \kappa + 9\gamma_1^2 + 4\delta^2 + \kappa^2 + 8\Omega^2} \\
&\times \sqrt{1+\frac{4\delta^2}{(\kappa+3\gamma_1)^2}} \cos(\phi - \mu) \bigg] \\
\mu =& -\arctan \Big(\frac{k+3\gamma_1}{2\delta}\Big)
	\end{split}
	\label{eq:phase_distr}
	\end{equation}
\begin{figure}
\subfloat[$\gamma_2/\gamma_1 = 1$]{%
  \includegraphics[width=0.499\linewidth]{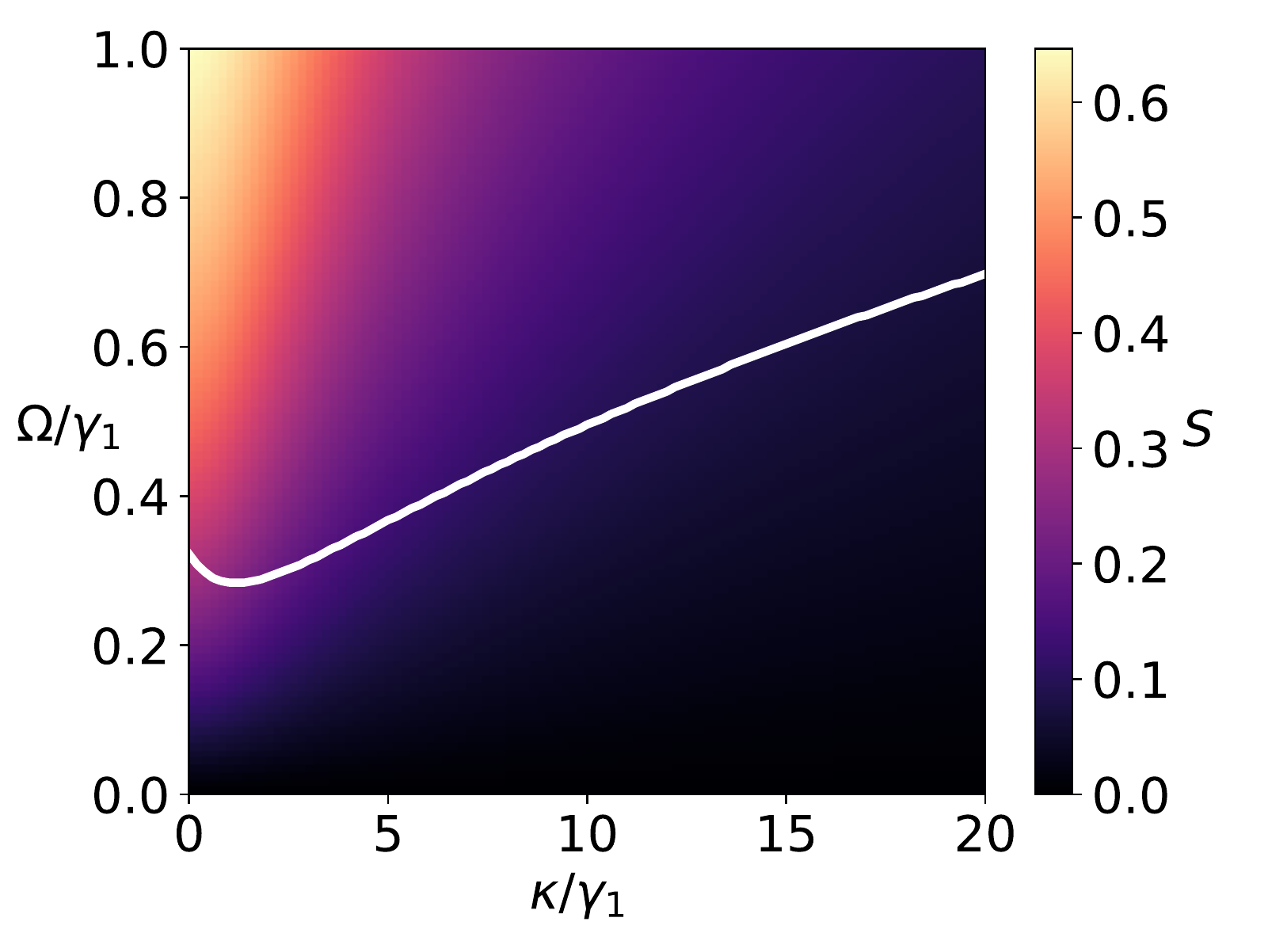}%
}
\hfill
\subfloat[$\gamma_2/\gamma_1 = 100$]{%
  \includegraphics[width=0.499\linewidth]{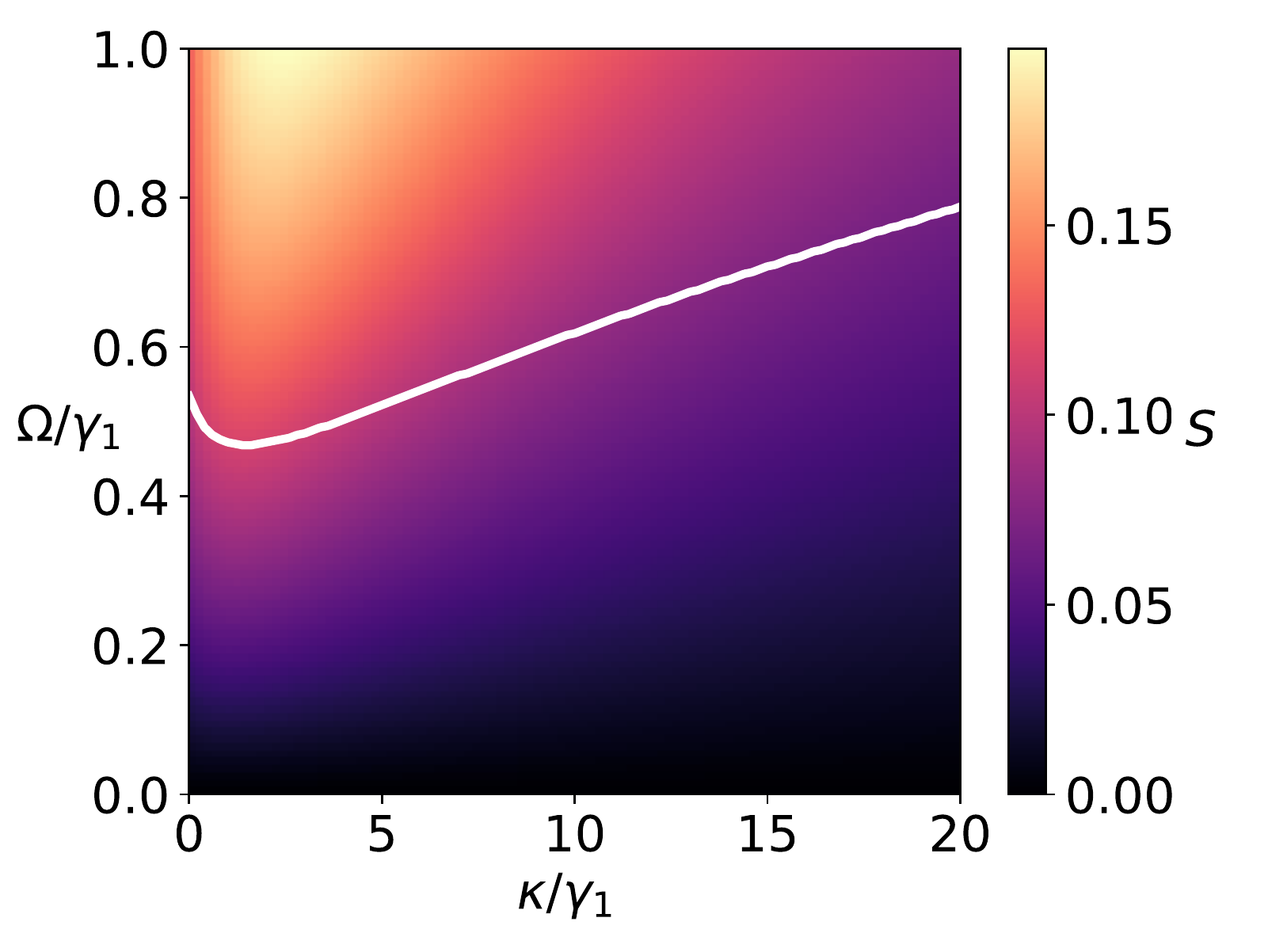}%
}
\hfill
	\caption{Synchronization as a function of driving strength $\Omega$ and noise parameter $\kappa$. The threshold driving ($\epsilon = 0.1$) is indicated by the white solid line.}
	\label{fig:S_plot}
	\end{figure}
It is a cardioid distribution \citep{mardia2009directional} $P(x) = \frac{1}{2\pi}(1-2S\cos(x-\mu))$ with MRL, $S$, and mean direction $\mu$. It is easy to see that the MRL of the distribution coincides with the sync measure in Eq.(\ref{eq:sync_dql}) calculated directly from the density matrix elements. This shows that the MRL, as statistically defined, can indeed be represented by the expectation value of a quantum operator. That enables easy numerical and analytical calculations of the amount of sync without projecting onto phase eigen-states and shows how synchronization is tied to the coherences of the density matrix. In this case, the expression for S gives an upper bound to synchronization by a harmonic drive in the deep quantum regime at $\lim_{\kappa, \Omega \to \infty} S = 1/(2\sqrt{2}) \approx 0.35$. This is unlike the behaviour of the (semi-)classical vdP oscillator\citep{kato2019semiclassical}, which approaches perfect synchronization ($S = 1$) for arbitrarily strong driving. The values that can be attained for moderate driving and noise are show in Fig.\ref{fig:S_plot}. The mean direction $\mu$, on the other hand, indicates the peak of the phase distribution. It is the analytical expression for the phase difference between the qvdP and the drive. As $\mu$ is given by an arctangent function, it always has a solution. For $\delta = 0$ the phase difference between the oscillator and drive will be 0, and for other detunings they will pick up a phase-difference in $[-\pi/2,\pi/2]$. Interestingly, the phase difference in the deep quantum regime is independent of the driving strength. This stands out from the classical\citep{balanov2009synchronization} and semi-classical regime\citep{kato2019semiclassical} where the synchronization phase depends on the driving strength. Classically, by driving harder, the phase difference eventually vanishes, whereas the quantum phase difference stays constant. The classical phase-difference also only has a solution in a range of detunings determined by the driving strength, whereas in the deep quantum regime it has a solution for arbitrary detuning. The expressions therefore set the qvdP in the deep quantum regime apart from classical oscillators, showing the robustness of quantum synchronization.

{\em Noise boosts synchronization.}-- In a physical implementation of a qvdP, some excitation relaxation of the oscillator (for example, as spontaneous emission of phonons in mechanical oscillators) is inevitable. This motivates considering non-negligible relaxation rates $\kappa \neq 0$. Adding linear damping in the deep quantum regime can, counter-intuitively, result in a boost in synchronization.

\begin{figure}
\subfloat[$\gamma_2/\gamma_1 = 1$]{%
  \includegraphics[width=0.5\linewidth]{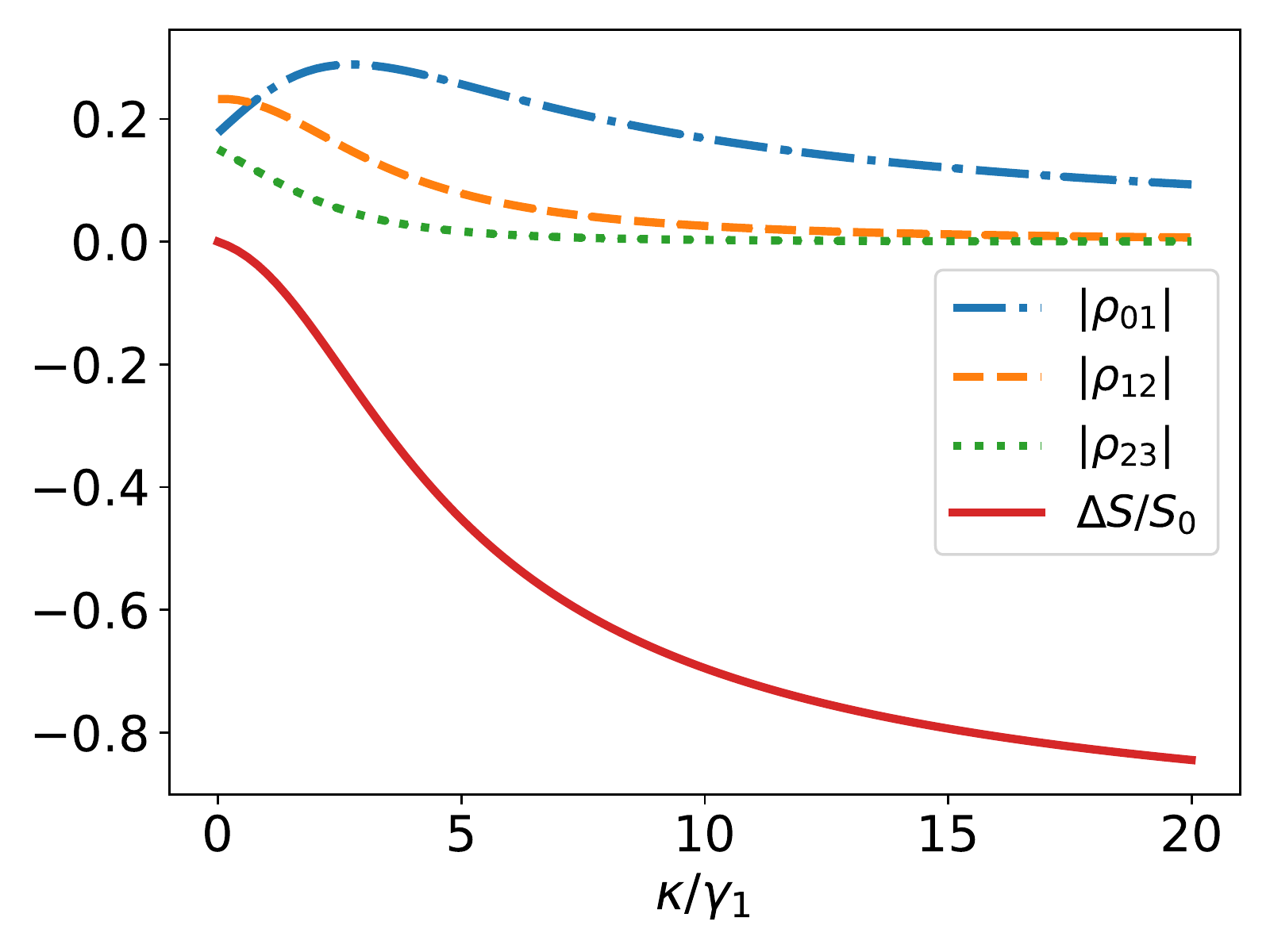}%
  \label{fig:coherences_b=1}%
}\hfill
\subfloat[$\gamma_2/\gamma_1 = 100$]{%
  \includegraphics[width=0.5\linewidth]{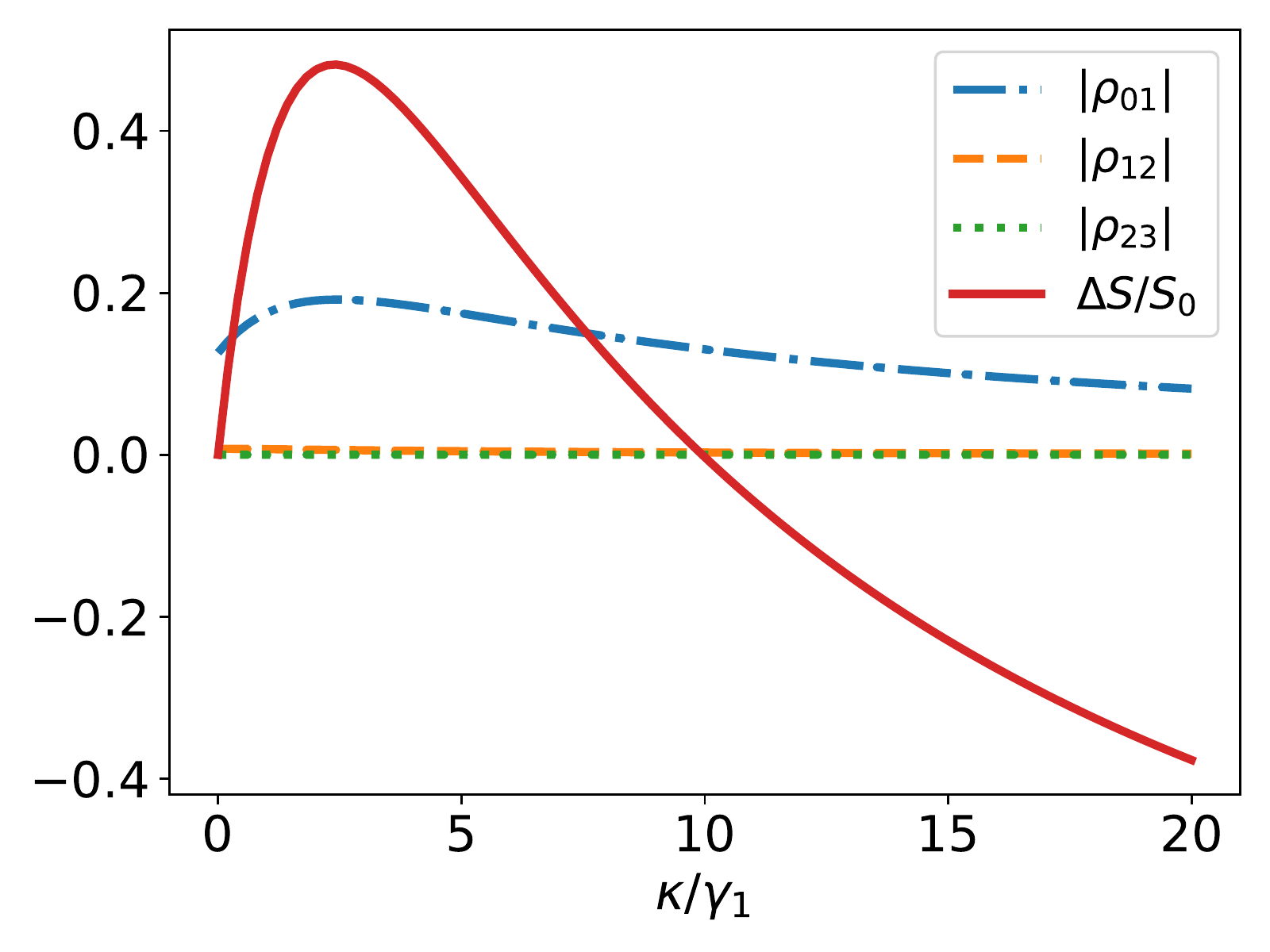}%
  \label{fig:coherences_b=100}%
}\hfill
\subfloat[Harmonic driving]{%
  \includegraphics[width=0.5\linewidth]{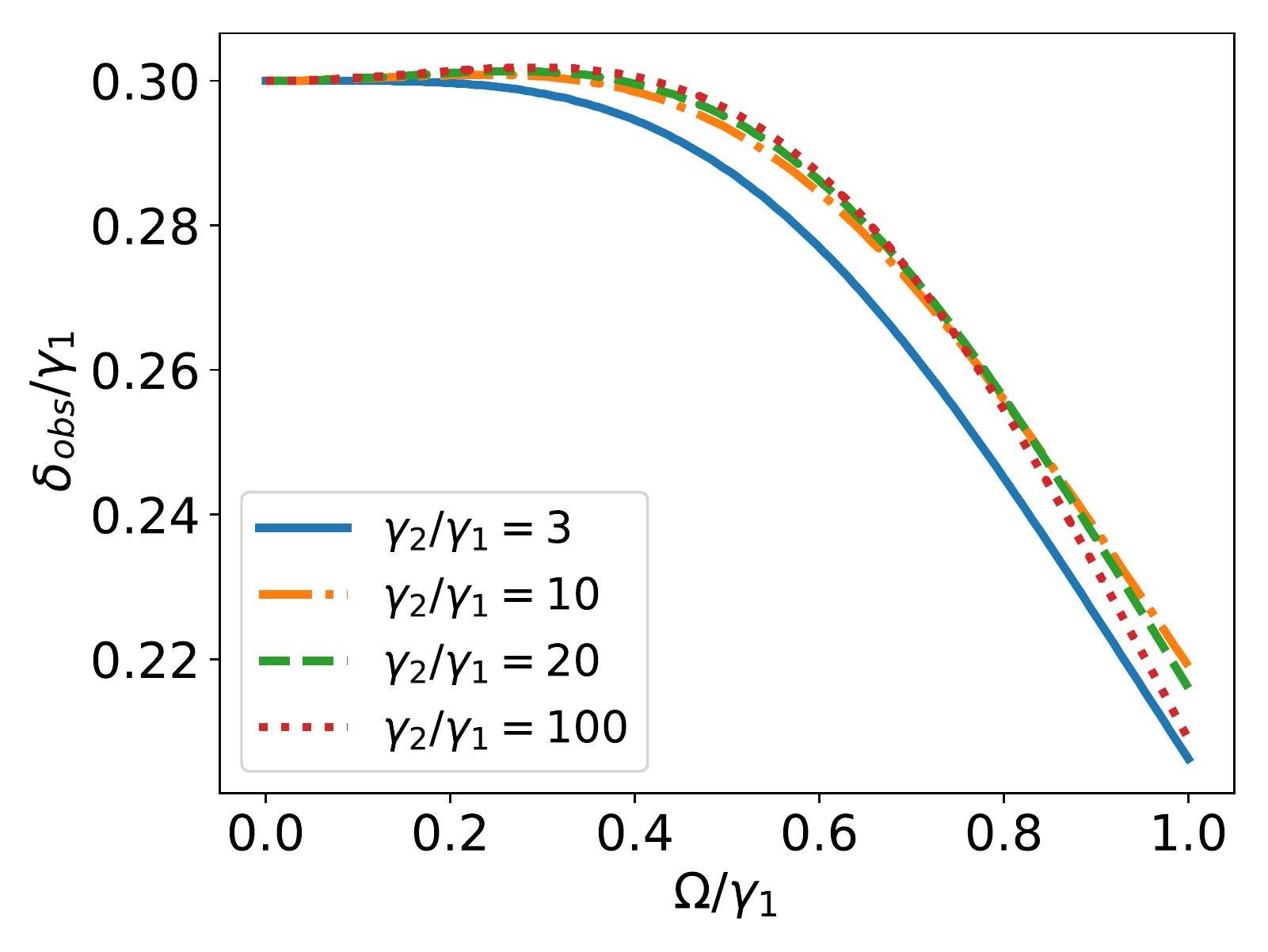}%
  \label{fig:entrainment_linear}%
}\hfill
\subfloat[Squeezed driving]{%
  \includegraphics[width=0.5\linewidth]{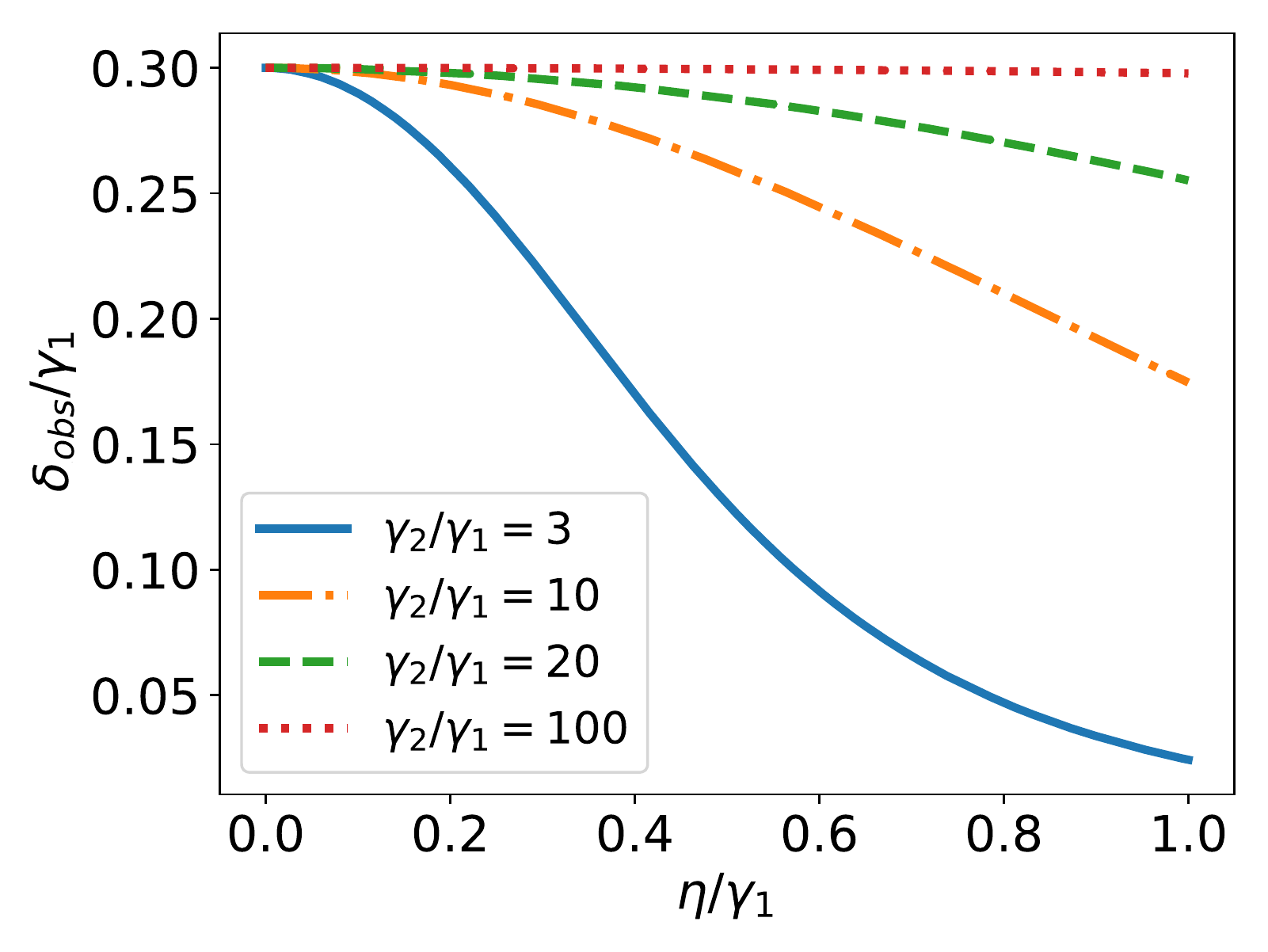}%
  \label{fig:entrainment_squeeze}%
}\hfill
\subfloat[Driving comparison]{%
  \includegraphics[width=0.5\linewidth]{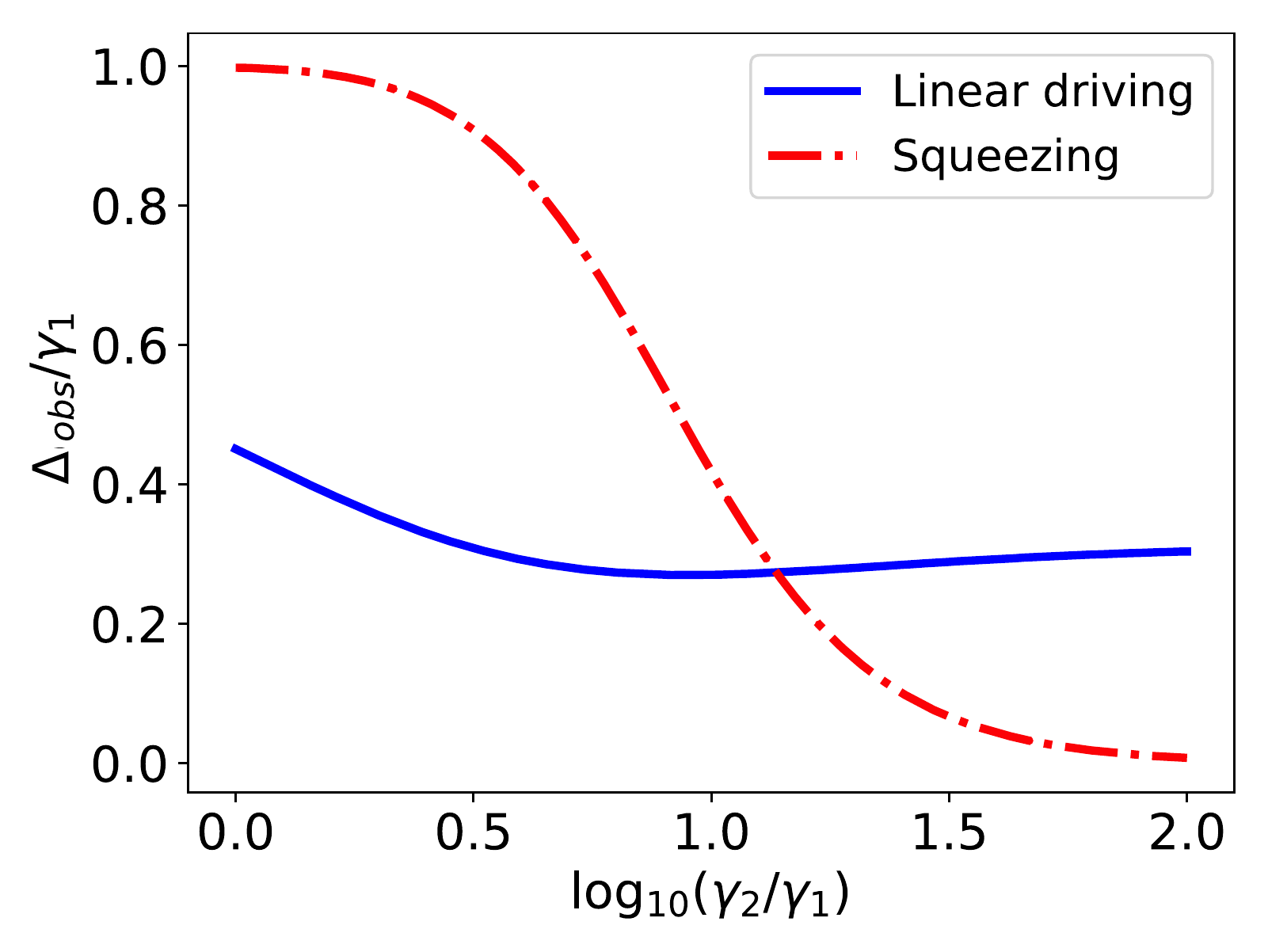}%
  \label{fig:entrainment_regime}%
}\hfill
\subfloat[Level diagram]{%
  \raisebox{-0.10cm}{\includegraphics[width=0.5\linewidth]{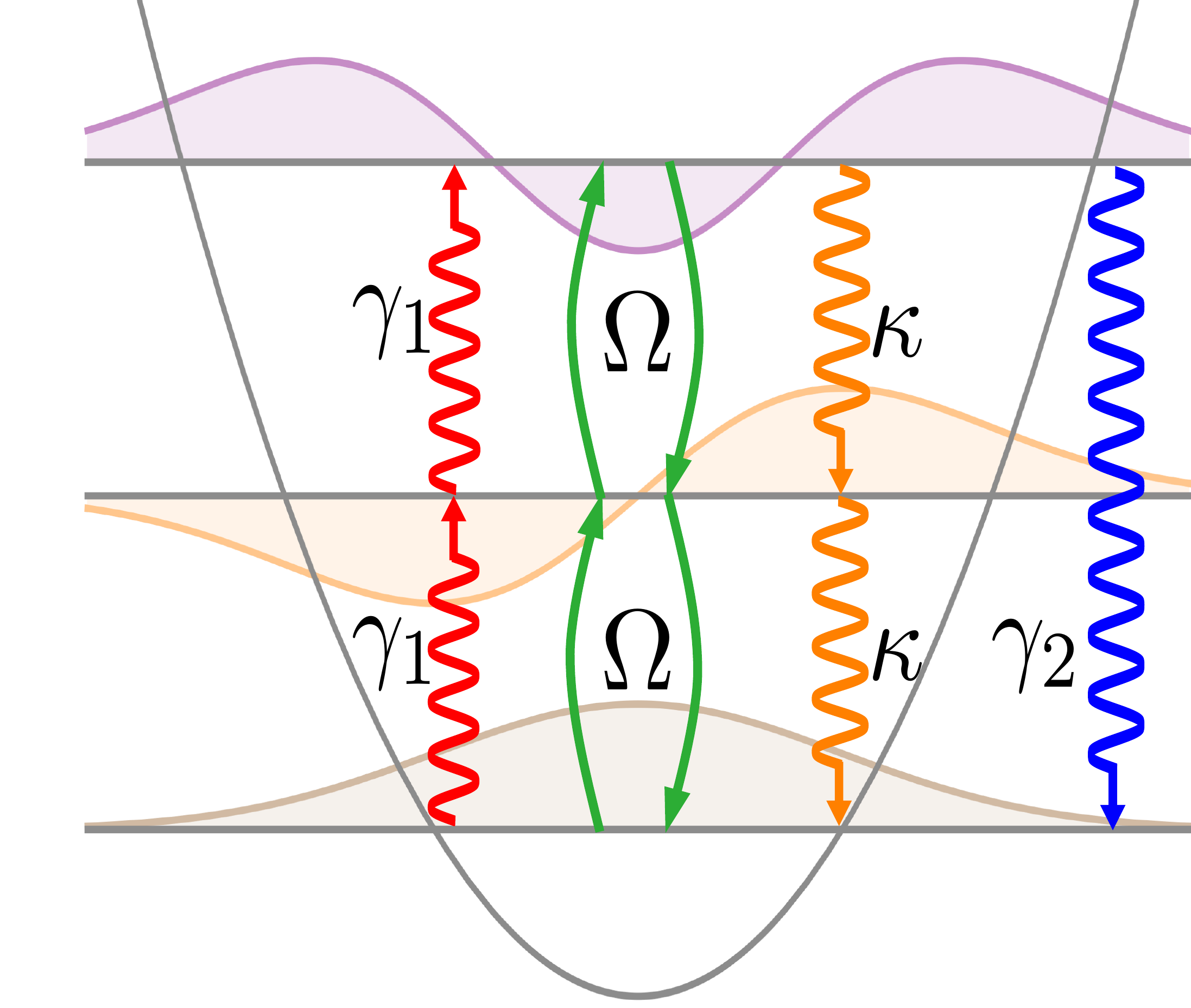}}%
  \label{fig:level_diagram}%
  }\hfill
  
	\caption{Effects of noise on coherences and synchronization for (a) $\gamma_2 / \gamma_1 = 1$ and (b) $\gamma_2 / \gamma_1 = 100$. (c)Level diagram of the qvdP in the deep quantum regime. $\gamma_1$ shows the effect of the single photon pumping, $\gamma_2$ the two-photon loss, $\kappa$ the single photon loss, and $\Omega$ the coherent driving.}
	\label{fig:coherences}

\end{figure}

Simply adding a non-zero $\kappa$ leads to a smaller limit cycle amplitude. To fairly compare synchronization of a noisy oscillator with a noise-free one, we must keep the relative driving strengths constant. This is achieved by always setting the driving strength such that it distorts the limit cycle by a constant amount $\epsilon$, such that $|\Delta N / N_0| = \epsilon$, where $\Delta N \equiv N - N_0$. We can then compare the synchronization measures of the two oscillators. In the deep quantum regime, the threshold driving strength is found using Eq.(\ref{eq:N_dql}): 
	\begin{equation}
	\label{eq:threshold}
\Omega_{\text{th}}^2 =\frac{\epsilon  \left(3 \gamma _1+\kappa \right) \left(6 \gamma _1 \kappa +9 \gamma _1^2+4 \delta ^2+\kappa ^2\right)}{4 \left[\gamma _1 (1-6 \epsilon )+\kappa (1-2  \epsilon) \right]}
	\end{equation}
Setting $\Omega = \Omega_{\text{th}}$ with resonant driving and varying $\kappa/\gamma_1$, we show numerically in Fig.\ref{fig:coherences} that the effect of noise differs dramatically in different regimes. The noise term $\kappa$ generally reduces coherences in the density matrix. The decoherence rate grows with the photon number, so higher elements in the density matrix decay faster, as seen in Fig.\subref*{fig:coherences_b=1} and \subref*{fig:coherences_b=100}. In the deep quantum limit the synchronization is given by only the lowest coherence $|\rho_{01}|$, which increases with the addition of noise. 

In Fig.\subref*{fig:level_diagram} we see that the coherences are all built by the driving $\Omega$. The element $|\rho_{01}|$ then depends on the probability of the oscillator being in in the state $\ket{0}$. In the absence of noise, once the oscillator has been driven into $\ket{1}$, it then then has to be driven to the state $\ket{2}$ before the driving can build up the coherence between the states $\ket{0}$ and $\ket{1}$. With weak driving, this is a very slow process. By adding the noise term, the oscillator can now decay down to the ground state $\ket{0}$ straight from $\ket{1}$. This increases the ground state population, which in turn increases the rate of transfer to $\ket{1}$. In the steady state, this shows up as a larger coherence $|\rho_{01}|$. Consequently, unique to the deep quantum regime, noise boosts synchronization. 

Not only is synchronization boosted in the deep quantum regime, but the limit-cycle is also extraordinarily robust to strong driving.  We can invert Eq.(\ref{eq:threshold}) and find that $\epsilon$ is bounded from above by a constant, independent of driving strength.
\begin{equation}
\label{eq:eps_bound}
\lim_{\Omega_{th} \rightarrow \infty}\epsilon \leq \frac{\gamma _1+\kappa }{2 \left(3 \gamma _1+\kappa \right)}
\end{equation}
Reducing in the noiseless limit to $\epsilon < 1/6$. This shows that quantum oscillators can be driven proportionally much harder than classical oscillators without breaking the assumption of self-sustained oscillations, further setting the qvdP apart from its classical counterpart. However, Eq.(\ref{eq:eps_bound}) show that the oscillator get less robust with the addition of noise. An unlimited driving strength is also still not allowed, as it will break the assumptions of the Lindblad-form master equation Eq.(\ref{eq:ME}), namely that the driving can be added perturbatively without influencing the derivation of the dissipators
\citep{shavit2019bridging}.

{\em Squeezing loses its effect.}-- Here, we consider the standard  noiseless ($\kappa = 0$) qvdP with pure squeeze driving. Squeezing was reported to enhance frequency entrainment at the damping ratio $\gamma_2 / \gamma_1 = 3$ \cite{sonar2018squeezing}. It was claimed to open up the possibility of observing quantum synchronization in the deep quantum regime. The intuition is that frequency (and phase) locking in the quantum regime is inhibited by quantum fluctuations \citep{walter2014quantum} which can be partially overcome by squeezing. The optimal combination of harmonic and squeeze driving has also been investigated \citep{koppenhofer2019optimal}. However, we will now show that the advantage of squeezing is diminished as the ratio $\gamma_2 / \gamma_1$ increases, and squeezing becomes totally ineffective in the deep quantum limit.

In Fig.\ref{fig:coherences}, the observed frequency (in the rotating frame of the driving) $\delta_{\text{obs}}$, which is the spectral peak frequency, is plotted against driving strength and damping ratio $\gamma_2 / \gamma_1$. In particular, Fig.\subref*{fig:entrainment_linear} shows that when the harmonic driving strength increases, the amount of frequency shift is relatively insensitive to the damping ratio. In contrast, Fig.\subref*{fig:entrainment_squeeze} shows that the entrainment due to squeezing decreases significantly as $\gamma_2 / \gamma_1$ increases, and becomes close to zero in the deep quantum regime. This difference in behaviour can be observed in greater detail in Fig.\subref*{fig:entrainment_regime}, where the relative entrainment $\Delta_{\text{obs}} = (\delta_{\text{obs}} - \delta) / \delta$ is plotted against the damping ratio, for a fixed driving strength. For harmonic driving, the entrainment does not depend greatly on the damping ratio. On the other hand, squeeze driving results in strong entrainment (near unity) for $\gamma_2 = \gamma_1$, but deteriorates to zero in the deep quantum limit. The crossover point where squeezing loses its advantage occurs at $\gamma_2 / \gamma_1 \approx 13$. Looking at the level-diagram in Fig.\subref*{fig:level_diagram} explains why squeeze driving loses its effect in the deep quantum regime. Squeezing is a two photon process, so it depends on the population of the state $\ket{2}$. In the deep quantum regime almost all the population is in the lowest two levels, so the squeezing drive decouples from the oscillator. This is another indication of new physics at $\gamma_2 / \gamma_1 \gtrsim 10$, motivating choosing it as the threshold for the deep quantum regime.

\section{Conclusion}

{\em Conclusion.}-- In this letter we have investigated the quantum van der Pol oscillator in the deep quantum regime. The regime was identified as $\gamma_2 /\gamma_1 \gtrsim 10$ from arguments based on where non-classical phenomena start to arise. We found that squeeze driving indeed provides a synchronization boost in the quantum regime, but in the deep quantum regime it totally loses its effect. Instead, we discovered a synchronization boost from noise, coming from a spontaneous decay of the oscillator into the environment. The effect is verified by exact numerical calculation of the master equation and by analytical approximations and explained through the dynamics of the coherences in the density matrix. This synchronization boost is another example of noise-enhanced processes, this time unique to the quantum domain. Furthermore, our synchronization measure enables us to show that there is a maximal amount of synchronization in the regime and that the oscillator can be driven much harder than its classical counterpart. We believe these results can act as fundamental building blocks for new phenomena in quantum nonlinear dynamics.

{\em Acknowledgements.}-- The authors would like to thank Tobias Haug for useful advice. The research is supported by the  Ministry of Education and the National Research Foundation, Singapore

\bibliographystyle{apsrev4-1}
\bibliography{qsync_bib}

\appendix
\newpage
\onecolumngrid

\section{Density matrix elements}

The full expressions for the steady-state density matrix elements are given by:
	\begin{equation}
\rho_{00} = \frac{1}{D} \bigg[ 2 \gamma _1 \left(\gamma _2 \left(4 \left(\delta ^2+\kappa ^2\right)+6 \Omega ^2\right)+3 \kappa  \left(\kappa ^2+2 \Omega ^2\right)\right)+\kappa  \left(\gamma _2+\kappa \right) \left(4 \delta ^2+\kappa ^2+4 \Omega ^2\right)+3 \gamma _1^2 \kappa  \left(7 \gamma _2+3 \kappa \right)+18 \gamma _2 \gamma _1^3 \bigg]
	\end{equation}
	\begin{equation}
\rho_{11} = \frac{1}{D} \bigg[\left(\gamma _2+\kappa \right) \left(\gamma _1 \left(6 \gamma _1 \kappa +9 \gamma _1^2+4 \delta ^2+\kappa ^2+12 \Omega ^2\right)+4 \kappa  \Omega ^2\right) \bigg]
	\end{equation}
	\begin{equation}
\rho_{22} = \frac{1}{D} \bigg[ \gamma _1 \left(\gamma _1 \left(6 \gamma _1 \kappa +9 \gamma _1^2+4 \delta ^2+\kappa ^2+12 \Omega ^2\right)+4 \kappa  \Omega ^2\right) \bigg]
	\end{equation}
	\begin{equation}
\rho_{01} = \frac{1}{D} \bigg[ - 2 \Omega  \left(\gamma _1 \left(\gamma _2-\kappa \right)+\kappa  \left(\gamma _2+\kappa \right)\right) \left(-3 i \gamma _1+2 \delta -i \kappa \right) \bigg]
	\end{equation}
where the denominator $D$ is
	\begin{equation}
	\begin{split}
D &= \gamma _1 \left[4 \gamma _1 \left(\delta ^2+4 \kappa ^2+3 \Omega ^2\right)+15 \gamma _1^2 \kappa +9 \gamma _1^3+4 \delta ^2 \kappa +7 \kappa  \left(\kappa ^2+4 \Omega ^2\right)\right] \\
&+ \gamma _2 \left(3 \gamma _1+\kappa \right) \left(6 \gamma _1 \kappa +9 \gamma _1^2+4 \delta ^2+\kappa ^2+8 \Omega ^2\right)+\kappa ^2 \left(4 \delta ^2+\kappa ^2+8 \Omega ^2\right)
	\end{split}
	\end{equation}
We can simplify the above expressions by taking the limit $\kappa \to 0$ (noiseless case). In this limit, the synchronization measure $S = | \rho_{01} |$ can be calculated to give
\begin{equation}
\lim_{\kappa/\gamma_1 \to 0} S = \frac{2 \gamma _2 \Omega  \sqrt{9 \gamma_1^2 + 4 \delta^2}}{4 \gamma _1 \left(\delta ^2+3 \Omega ^2\right)+3 \gamma _2 \left(9 \gamma _1^2+4 \delta ^2+8 \Omega ^2\right)+9 \gamma _1^3}
	\label{eq:sync_k=0}
	\end{equation}
which can be used to describe synchronization of the noiseless quantum van der Pol (qvdP) oscillator in the quantum regime (without going into the deep quantum regime). On the other hand, we can also consider the noisy qvdP in the deep quantum limit, for which the synchronization measure becomes
	\begin{equation}
\lim_{\gamma_2/\gamma_1 \to \infty} S= \frac{2 \Omega  \left(\gamma _1+\kappa \right)\sqrt{(3 \gamma_1 + \kappa)^2 + 4\delta^2}}{\left(3 \gamma _1+\kappa \right) \left(6 \gamma _1 \kappa +9 \gamma _1^2+4 \delta ^2+\kappa ^2+8 \Omega ^2\right)}
	\label{eq:sync_dql_appendix}
	\end{equation}
	
\section{Synchronization boost due to noise}
An intuitive picture to explain the synchronization boost is provided in the main text (in terms of the effects of noise on the various coherences). One can also observe the boost by inspecting Eq.(\ref{eq:sync_dql_appendix}). In the deep quantum limit, since	
	\begin{equation}
\frac{\partial S}{\partial \kappa}\bigg|_{\kappa = 0} = \frac{2 \Omega  \left(3 \gamma _1^2+8 \Omega ^2\right)}{\left(9 \gamma _1^2+8 \Omega ^2\right)^2} > 0
	\label{eq:ds_dkappa}
	\end{equation}
for any fixed $\Omega > 0$, there will still be an initial synchronization boost when increasing $\kappa$ from zero even after discounting the fact that $\Omega_{\text{th}}$ increases with $\kappa$. At the very least, we can conclude that synchronization in the deep quantum regime is highly robust against relaxation losses, a feature not present in the more classical case. This effect, of course, is not limited to only the deep quantum limit. Using the full expression of the density matrix, we can also impose the condition $\partial_\kappa S (\kappa = 0) > 0$. To this end, we start with the synchronization measure (without taking limits):
	\begin{equation}
S = |\rho_{01}| = \frac{M}{D}
	\end{equation}
where
	\begin{equation}
M = 2 \Omega [\gamma_1 (\gamma_2 - \kappa) + \kappa (\gamma_2 + \kappa)] \sqrt{4\delta^2 + (\kappa + 3\gamma_1)^2}
	\end{equation}
We assume $\gamma_2 > \kappa$ such that $M > 0$. The condition $\partial_\kappa S (\kappa = 0) > 0$ is thus equivalent to (with $M^\prime \equiv \partial_\kappa M (\kappa = 0)$ and similarly for $D^\prime$)
	\begin{equation}
\frac{M^\prime}{M} > \frac{D^\prime}{D}
	\end{equation}
Noting that both $D$ and $D^\prime$ are positive, a necessary condition for synchronization boost is therefore $M^\prime / M > 0$. Substituting the expressions for $M^\prime$ and $M$ (evaluated at $\kappa = 0$), the necessary condition is
	\begin{equation}
\frac{\gamma_2}{\gamma_1} > 1 - \frac{3}{4(\delta/ \gamma_1)^2 + 12}
	\end{equation}
In other words, if the damping ratio $\gamma_2 / \gamma_1 < 3/4$ (corresponding to resonant driving), it is impossible for noise to enhance synchronization. This shows that synchronization boost due to noise is a purely quantum effect. This also aligns with our physical picture that the enhancement originates from the boost of the lowest coherence $|\rho_{01}|$, which has no classical analogue.

\section{Comparison of the analytical and numerical Arnold tongues}
Using Eq. (\ref{eq:sync_k=0}), the synchronization measure can be plotted against detuning and driving strength, giving the so-called Arnold tongue, which is a key signature of synchronization. The damping ratio $\gamma_2 / \gamma_1$ is set as 100 in order to reach the deep quantum regime. The analytical solution agrees with the numerical simulation at low levels of driving. For stronger driving however, the analytical solution begins to differ from numerical results. This may be due to two reasons: (1) the coherence $\rho_{13}$ and/or $\rho_{23}$ becomes non-negligible due to the driving, which thus breaks the assumption made in our solution, or (2) the external driving causes a significant distortion in the limit cycle, which contradicts the very definition of synchronization which requires the driving to only be a perturbation to the limit cycle. 

To ensure that the limit cycle is not greatly distorted, one could set a threshold $\epsilon$ to impose the condition for synchronization that $|\Delta N / N_0| < \epsilon$, where $\Delta N \equiv N - N_0$. We set the maximum allowed distortion to be $\epsilon = 0.1$. Note that while the choice of $\epsilon$ is arbitrary, $\epsilon$ has to be suitably small to prevent large distortions to the limit cycle. 
	\begin{figure}
\subfloat{%
  \includegraphics[width=0.499\linewidth]{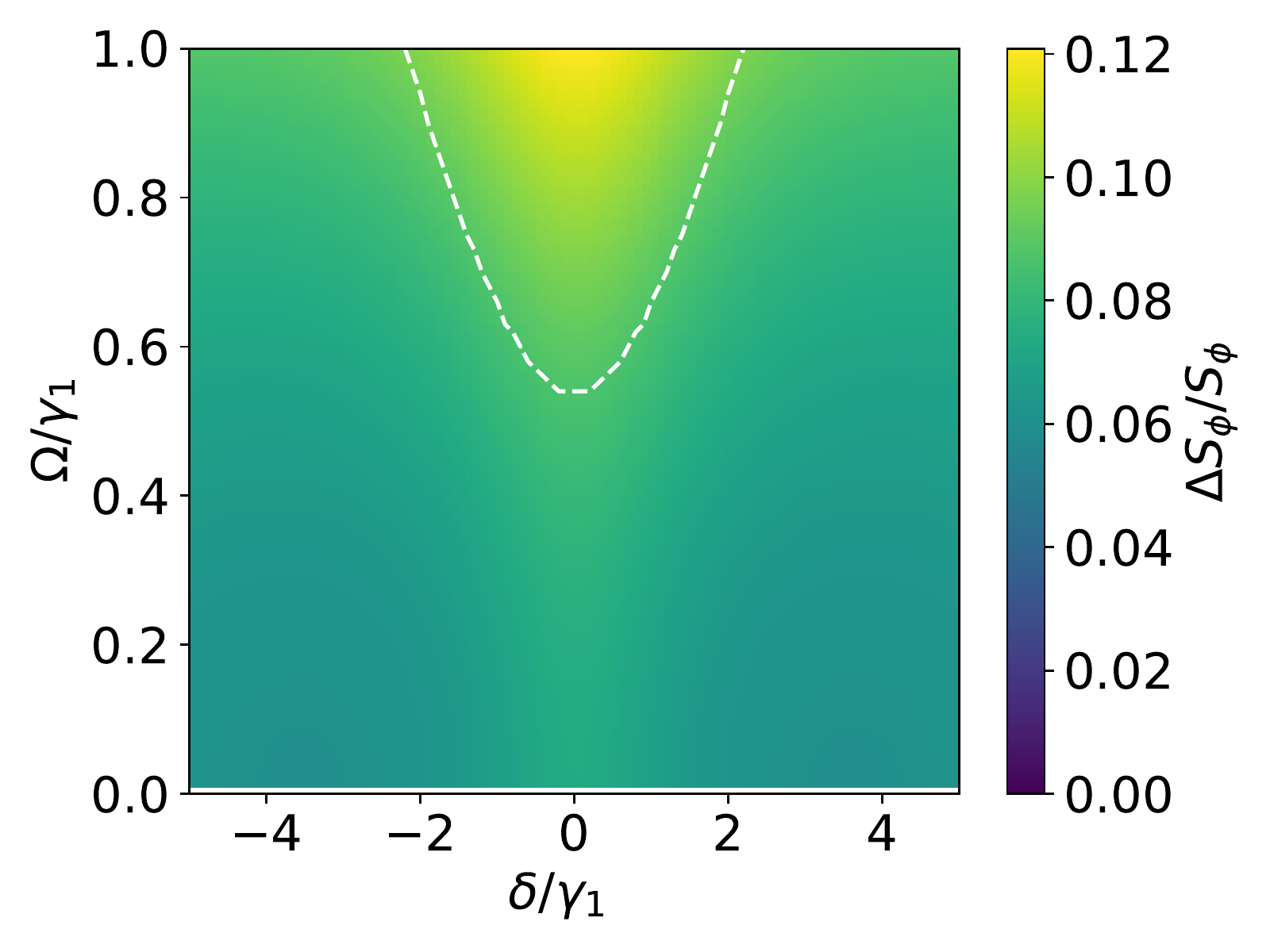}%
  \label{fig:arnold_diff}%
}\hfill
\subfloat{%
  \includegraphics[width=0.499\linewidth]{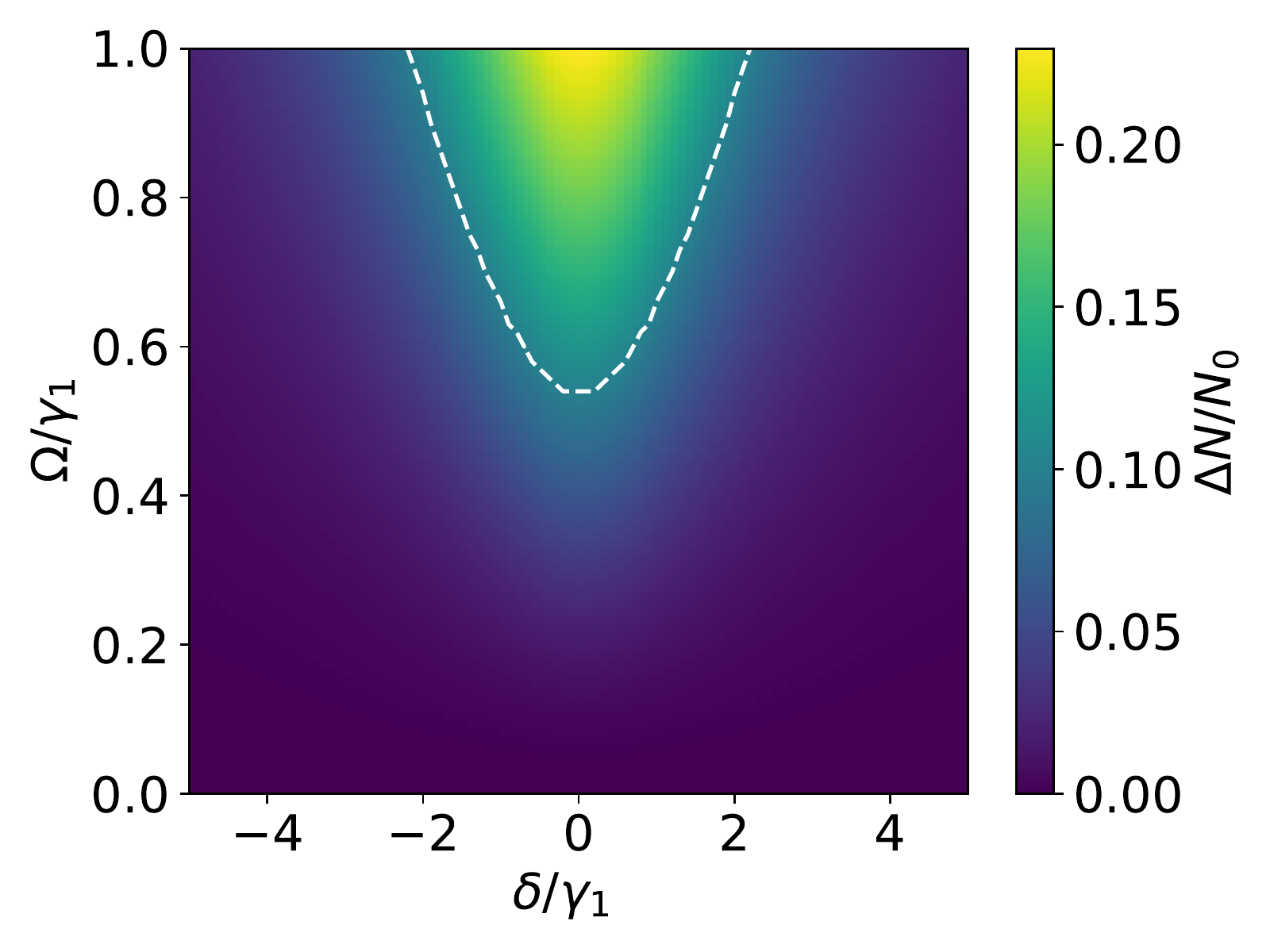}%
  \label{fig:arnold_distortion}%
}
\hfill
\subfloat{%
  \includegraphics[width=0.499\linewidth]{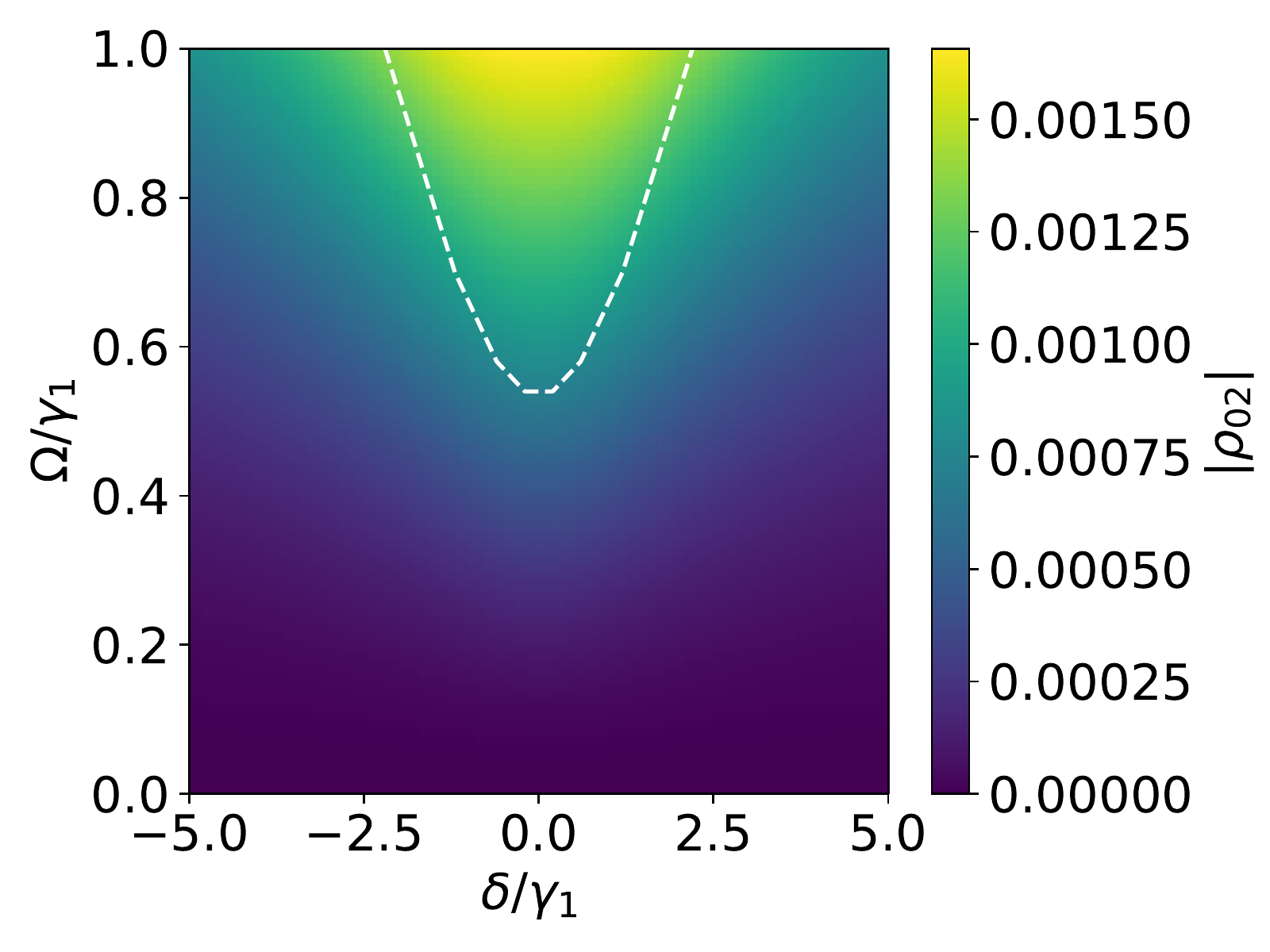}%
  \label{fig:coh_p02}%
}\hfill
\subfloat{%
  \includegraphics[width=0.499\linewidth]{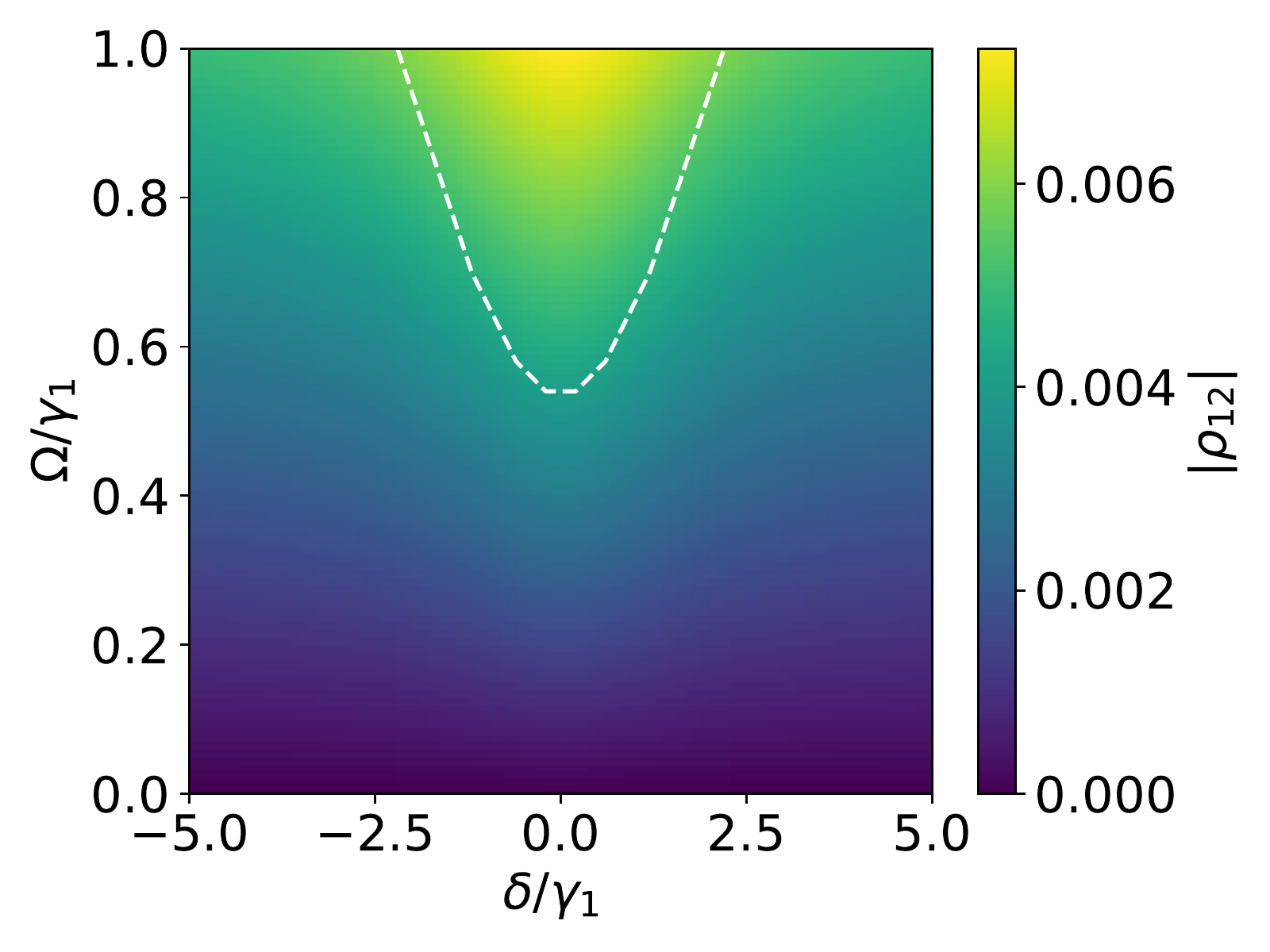}%
  \label{fig:coh_p12}%
}
\hfill
	\caption{(a) Difference of the Arnold tongue between numerical simulation and analytical solution (b) Distortion of limit cycle, measured by the change in amplitude $\Delta N \equiv N - N_0$. (c) The magnitude of the coherence $|\rho_{02}| = |\braket{0|\rho|2}|$. (d) The magnitude of the coherence $|\rho_{12}| = |\braket{1|\rho|2}|$. The white dashed line marks the threshold $\epsilon = 0.1$. The damping ratio is set at $\gamma_2 / \gamma_1 = 100$.}
	\label{fig:arnold_threshold}
	\end{figure}	
To quantify the accuracy of our analytical solution, we compute the difference between the synchronization measures, as shown in Fig. \subref*{fig:arnold_diff}. Generally, using the threshold of $\epsilon = 0.1$, the analytical solution is within around 10\% accuracy of the the numerical simulation. The region with the most significant difference occurs where $\delta = 0$ and relatively stronger driving, causing as high as 12\% difference in the result. However, the region which exceeds the distortion threshold (marked out by the white dashed line) should not be regarded as synchronization due to the significant distortion of the limit cycle caused by driving.

For a better understanding of the slight inaccuracy of the analytical solution, we plot the magnitude of the coherences $|\rho_{02}| = |\braket{0|\rho|2}|$ and $|\rho_{12}| = |\braket{0|\rho|2}|$ in Fig. \subref*{fig:coh_p02} and Fig. \subref*{fig:coh_p12} respectively, where $\rho$ is the steady state density matrix. Unsurprisingly, both $|\rho_{02}|$ and $| \rho_{12}|$ are higher in the `disallowed' region  exceeding the threshold $\epsilon = 0.1$, primarily due to the stronger driving. It should also be noted however that $|\rho_{12}| > |\rho_{02}|$. Again, this is not surprising because $\rho_{02}$ relies on two-photon process which is second-order in $\Omega$, thus explaining the small contribution when considering relatively weak driving.

\end{document}